\documentclass[a4paper]{jpconf}
\usepackage{graphicx}
\usepackage{amsfonts}
\usepackage{amsmath}
\usepackage{amssymb}
\usepackage{latexsym}

\newcommand{\BbbR}{\mathbb{R}}
\newcommand{\BbbZ}{\mathbb{Z}}
\newcommand{\RPthree}{\mathbb{RP}^3}
\newcommand{\RPtwo}{\mathbb{RP}^2}

\numberwithin{equation}{section}

\begin{document}
\title{Geon black holes and quantum 
field theory}

\author{Jorma Louko}

\address{School of Mathematical Sciences, 
University of Nottingham, 
Nottingham NG7~2RD, UK}

\ead{jorma.louko@nottingham.ac.uk}

\begin{abstract}
Black hole spacetimes that are 
topological geons in the sense of Sorkin can 
be constructed by taking a quotient 
of a stationary black hole that has a bifurcate 
Killing horizon. We discuss the geometric properties 
of these geon black holes and the Hawking-Unruh effect 
on them. We in particular show how correlations in the 
Hawking-Unruh effect reveal to an exterior observer 
features of the geometry that are classically confined 
to the regions behind the horizons. 

$\phantom{xxx}$ 

\noindent 
Talk given at the 
{\sl First Mediterranean Conference on
Classical and Quantum Gravity\/}, 
Kolymbari (Crete, Greece), 
September 14--18, 2009. 

$\phantom{xxx}$ 

\noindent 
\emph{This contribution is dedicated to Rafael Sorkin
in celebration of his influence, 
inspiration and friendship.} 

\end{abstract}

\section{Introduction}
A geon, short for 
``gravitational-electromagnetic entity'',  
was introduced in 1955 by 
John Archibald Wheeler \cite{wheeler-geon}  
as a configuration of the gravitational 
field, possibly coupled to other zero-mass fields
such as massless neutrinos \cite{brill-wheeler-geon} or the 
electromagnetic field~\cite{brill-hartle-geon},  
such that a distant observer sees the curvature 
to be concentrated in a central region 
with persistent large scale features. 
The configuration was required to be 
asymptotically flat~\cite{melvin-prd}, 
allowing the mass to be defined by what are now known as 
Arnowitt-Deser-Misner 
% (ADM) 
methods. 
The examples discussed in \cite{MW-geondata} 
indicate that the configuration was also 
understood to have spatial topology~$\BbbR^3$, 
excluding black holes. 
These geons are however believed to be unstable, owing to 
the tendency of massless fields either 
to disperse to infinity or to collapse into a 
black hole~\cite{gundlach-rev}.

% One might say: ``Nice idea, did not quite work.'' 

\begin{figure}[b]
\includegraphics[width=24pc]{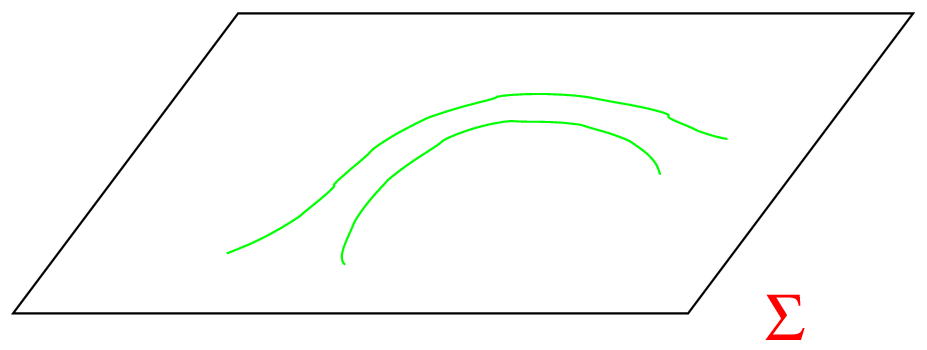}\hspace{2pc}%
\begin{minipage}[b]{10pc}\caption{\label{fig:topogeon}A spatial 
slice $\Sigma$ of a topological geon. The geometry of $\Sigma$ is 
asymptotically Euclidean, but the topology of $\Sigma$ 
does not need to be~$\BbbR^3$, as illustrated by the handle drawn.}
\end{minipage}
\end{figure}

In 1985, Sorkin \cite{sorkin-topogeon}
generalised Wheeler's geon into a \emph{topological geon\/}
by allowing the spatial topology to be nontrivial, 
as well as dropping the a priori requirement of 
persistent large scale features. 
The central regions may thus have complicated topology, 
as illustrated in Figure~\ref{fig:topogeon},  
and the time evolution of the initial data may lead 
into singularities and black holes~\cite{gannon1,gannon2}. 
In particular, topological geons include quotients of 
stationary black hole 
spacetimes with a bifurcate Killing horizon, 
such that the two exteriors separated by 
the Killing horizon become identified 
\cite{FriedSchWi,Louko:2004ej}. 
Such geon black holes are an intermediate case between 
conventional stationary black holes and dynamical black holes, 
in the sense that the nonstationary features are 
confined behind the horizons. 

As the nontrivial spatial topology in a geon 
black hole is present since arbitrarily early times, 
one does not expect an astrophysical star 
collapse to result into a geon black hole. 
Instead, the interest of geon black holes is in their 
quantum mechanical properties. 
Because the Killing horizon now does not separate 
two causally disconnected exteriors, 
it is not possible to arrive at a thermal density matrix 
for a quantum field 
by the usual procedure of tracing 
over the second exterior. 
Nevertheless, as we shall discuss, 
a quantum field on a geon black hole 
does exhibit thermality in the standard Hawking 
temperature, although only for a restricted set of observations. 
We shall in particular see how the quantum 
correlations in the Hawking-Unruh effect 
reveal to an exterior observer features of the 
geometry that are confined to the regions 
behind the horizons. 

% The rest of the paper is as follows. 

We start by introducing in Section 
\ref{sec:showcase}
the showcase example, the $\RPthree$ geon, 
formed as a quotient of Kruskal. 
Section 
\ref{sec:othergeons}
discusses a number of generalisations, 
including geons with angular momenta and gauge charges. 
Thermal effects in quantum field 
theory on geons are addressed in Section~\ref{sec:qft}. 
Section \ref{sec:conclusions} 
concludes by discussing the prospects of 
computing the entropy of a geon 
from a quantum theory of gravity.

\section{Showcase example: $\RPthree$ geon}
\label{sec:showcase}

\begin{figure}[p]
\begin{minipage}{18pc}
\includegraphics[width=18pc]{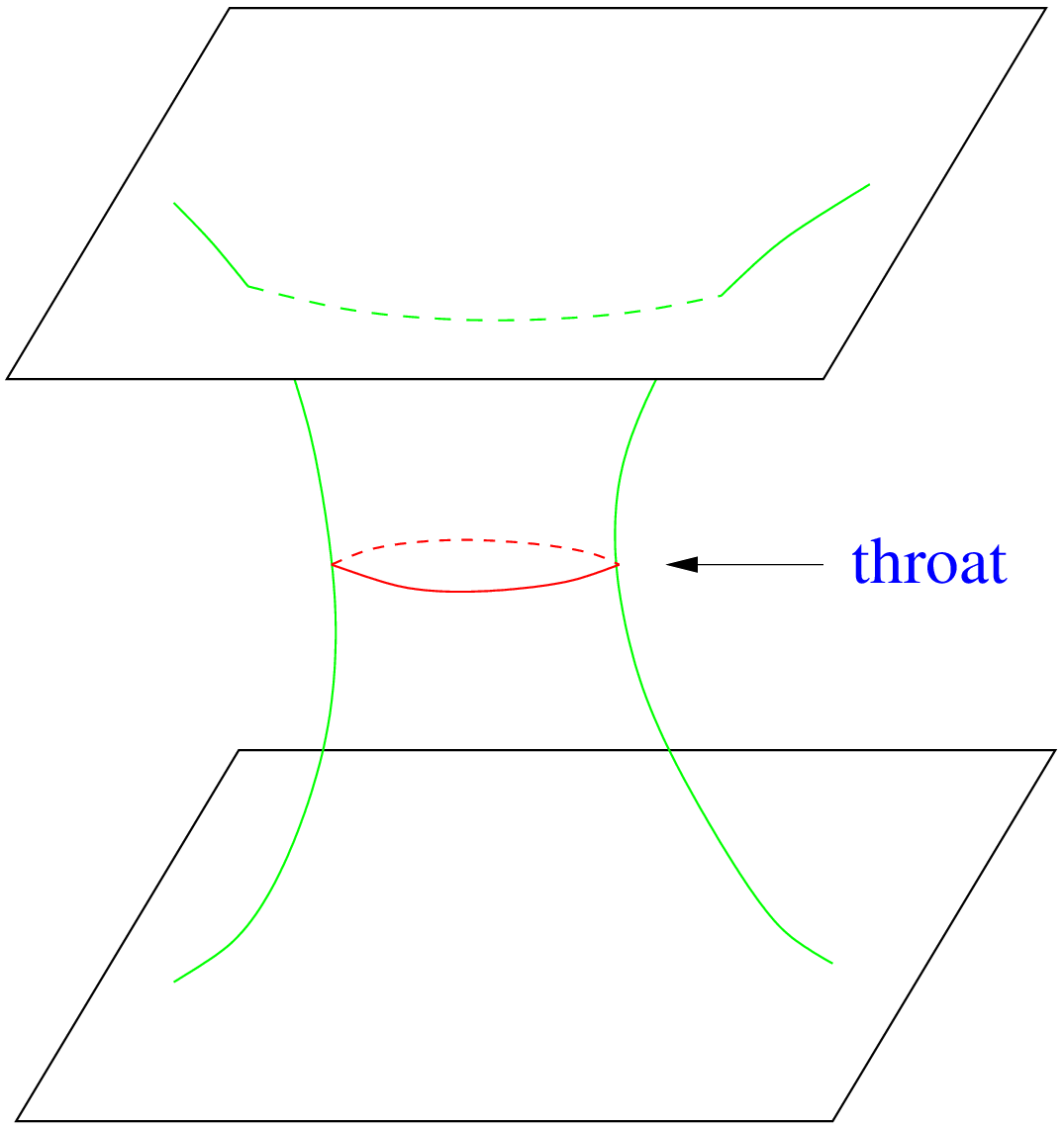}
\caption{\label{fig:wormhole}Time-symmetric wormhole initial 
data for Kruskal manifold, with the radial dimension 
drawn vertical and the two-spheres 
of spherical symmetry drawn as
horizontal circles. 
The wormhole 
connects two asymptotically flat infinities. 
It admits a 
freely-acting involutive isometry that consists of the 
radial reflection about the wormhole throat 
composed with the antipodal map on the~$S^2$.}
\end{minipage}\hspace{2pc}%
\begin{minipage}{18pc}
\includegraphics[width=18pc]{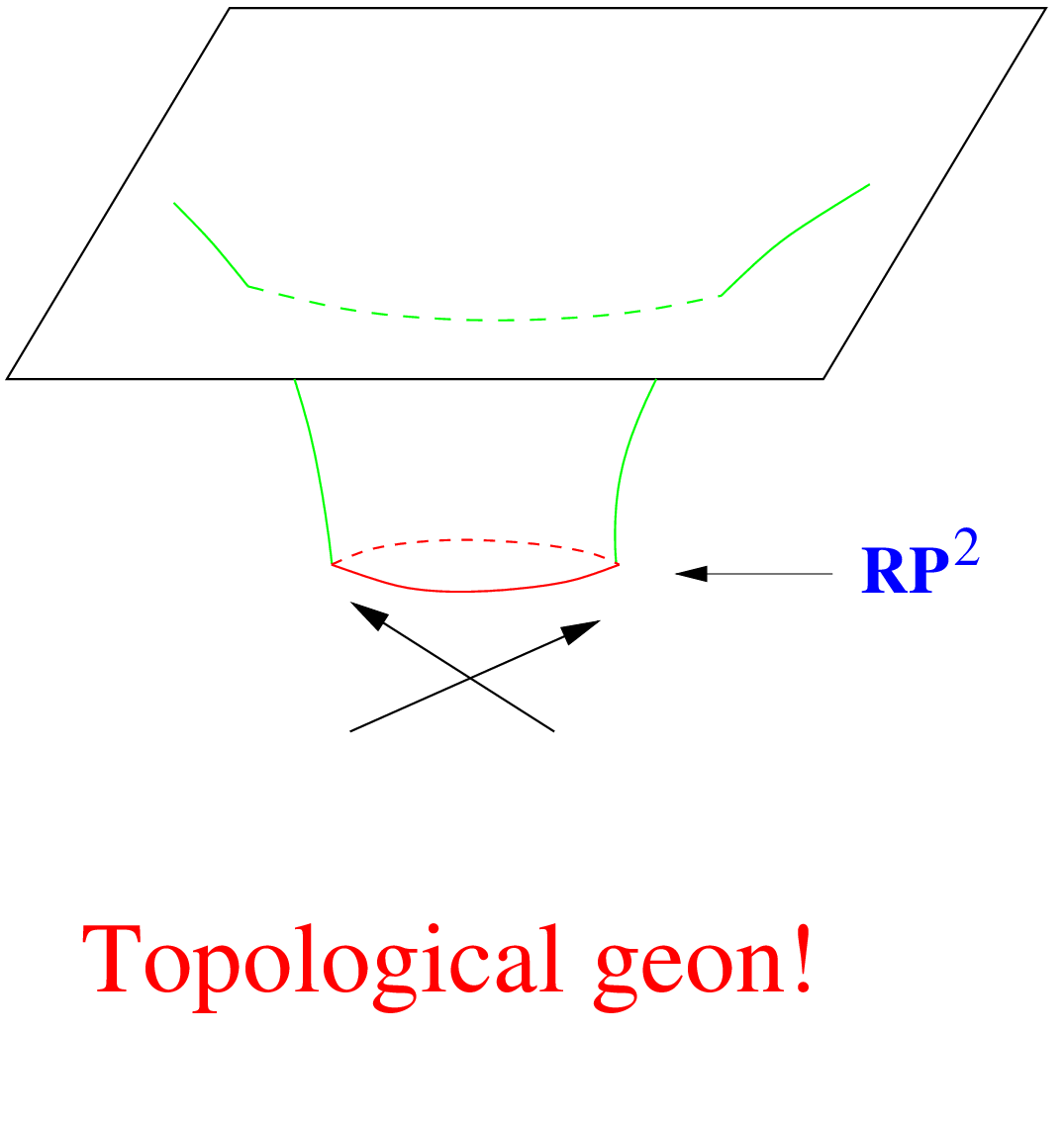}
\caption{\label{fig:wormholez2}Time-symmetric initial data for 
the $\RPthree$ geon, obtained as the $\BbbZ_2$ quotient of the 
wormhole by the freely-acting involutive isometry. 
The topology of the wormhole is 
$S^2 \times \BbbR \simeq S^3 \setminus 
\{\text{2 points}\}$
and that of the quotient is 
$\bigl(S^3 \setminus 
\{\text{2 points}\}\bigr)/\mathbb{Z}_2 \simeq
\RPthree \setminus \{\text{point}\}$, 
each of the omitted points being at 
an asymptotically flat infinity.\\
$\phantom{xxx}$\\
$\phantom{xxx}$
}
\end{minipage} 
\end{figure}

Before the Kruskal-Szekeres extension 
of the Schwarzschild solution was known, it 
had been observed that the 
time-symmetric initial data for exterior Schwarzschild 
can be extended into a time-symmetric wormhole initial 
data, connecting two asymptotically flat infinities, 
illustrated in Figure~\ref{fig:wormhole}. 
Misner and Wheeler \cite{MW-geondata} noted that 
this wormhole admits an involutive freely-acting isometry 
that consists of the radial reflection about the wormhole 
throat composed with the antipodal map on 
the $S^2$ of spherical symmetry. 
The $\BbbZ_2$ quotient of the wormhole by this isometry, 
illustrated in 
Figure~\ref{fig:wormholez2}, is hence a new 
time-symmetric spherically symmetric initial 
data for an Einstein spacetime. 

\begin{figure}[p]
\begin{minipage}{18pc}
\includegraphics[height=10pc]{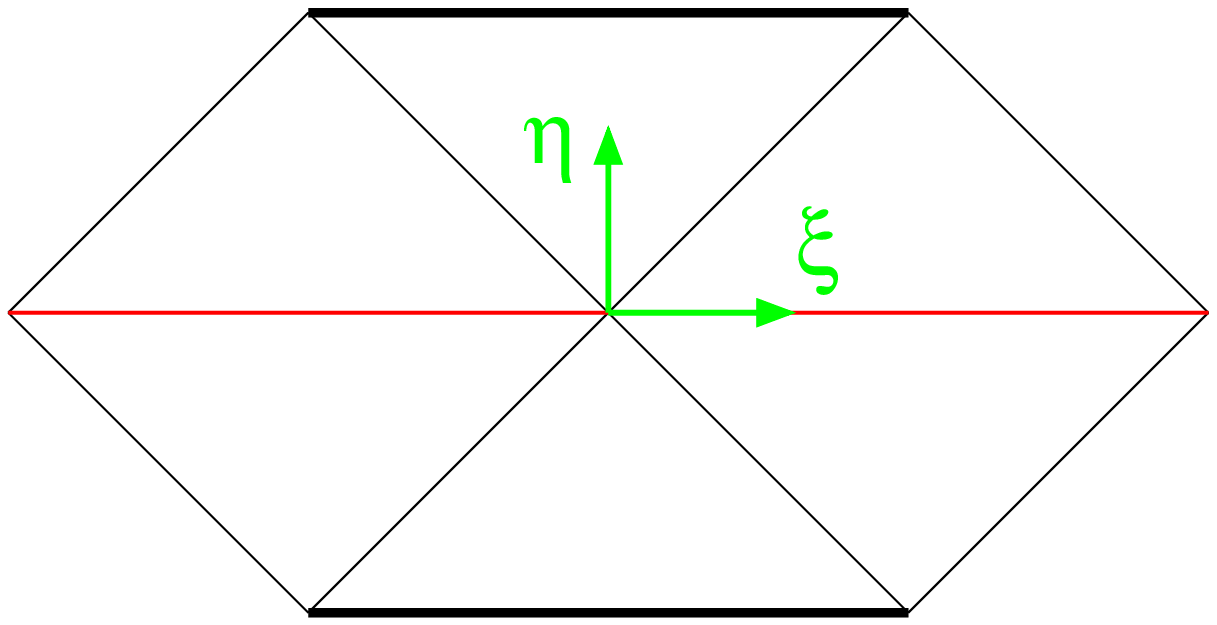}
\caption{\label{fig:kruskal}Conformal diagram of 
Kruskal manifold, 
with the $S^2$ orbits of spherical symmetry suppressed and 
the Kruskal time and space coordinates $(\eta,\xi)$ 
shown. The wormhole of Figure \ref{fig:wormhole} is the horizontal 
line $\eta=0$. 
The freely-acting involutive isometry that 
preserves time orientation and restricts to that of the wormhole is 
$J_K: \bigl(
\eta,
\xi, \theta, \varphi\bigr) \mapsto 
\bigl(
\eta , - 
\xi , 
P 
(\theta, \varphi)\bigr)$, where $P$ is the $S^2$ antipodal map.}
\end{minipage}\hspace{2pc}%
\begin{minipage}{18pc}
\hspace{2pc}\includegraphics[height=10pc]{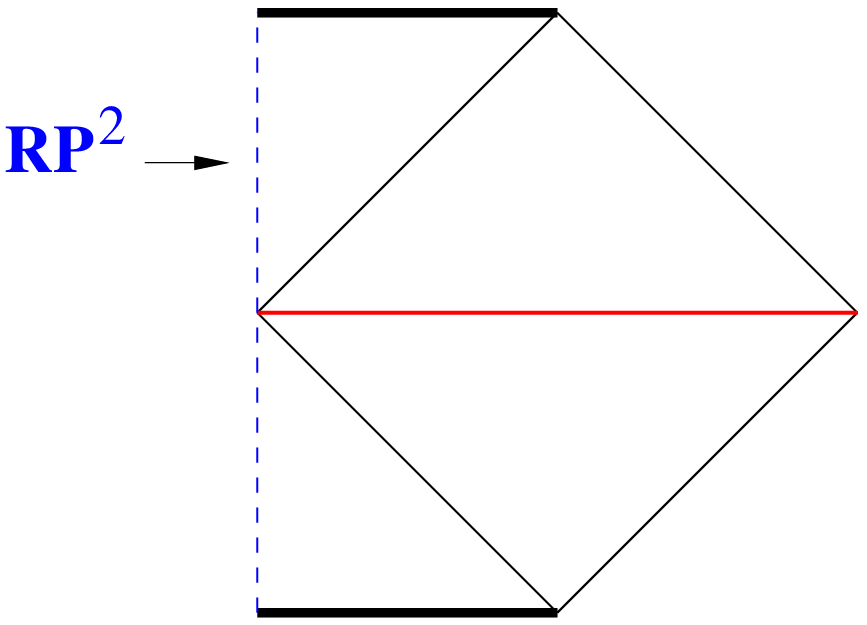}
\caption{\label{fig:rp3geon}Conformal diagram of 
the $\RPthree$ geon, obtained as the $\BbbZ_2$ 
quotient of Kruskal by the freely-acting 
involutive isometry~$J_K$. 
The suppressed orbits of spherical symmetry are $S^2$ 
except on the dashed 
line where they are~$\RPtwo$.
The initial data of Figure \ref{fig:wormholez2} 
is shown as the horizontal line.\\
$\phantom{xxx}$\\
$\phantom{xxx}$\\
$\phantom{xxx}$}
\end{minipage} 
\end{figure}

The new time-symmetric initial data has topology 
$\RPthree \setminus \{\text{point}\}$, 
the omitted point being at 
the asymptotically flat infinity~\cite{giulini-thesis}. 
The time evolution is a topological geon spacetime, 
called the 
$\RPthree$ geon~\cite{FriedSchWi}, 
and it can be obtained as a 
$\BbbZ_2$ quotient of 
Kruskal as shown in Figures 
\ref{fig:kruskal}
and~\ref{fig:rp3geon}. 
A~general discussion of quotients of Kruskal 
can be found in~\cite{chamblin-gibbons}. 

The $\RPthree$ geon is a black and white hole spacetime, 
it is spherically symmetric, and it is time and space orientable. 
It inherits from Kruskal a black hole singularity 
and a white hole singularity, 
but the quotient has introduced no new singularities. 
The distinctive feature is that is has only one exterior region, 
isometric to exterior Schwarzschild. 

\begin{figure}[t]
\begin{minipage}{18pc}
\includegraphics[height=10pc]{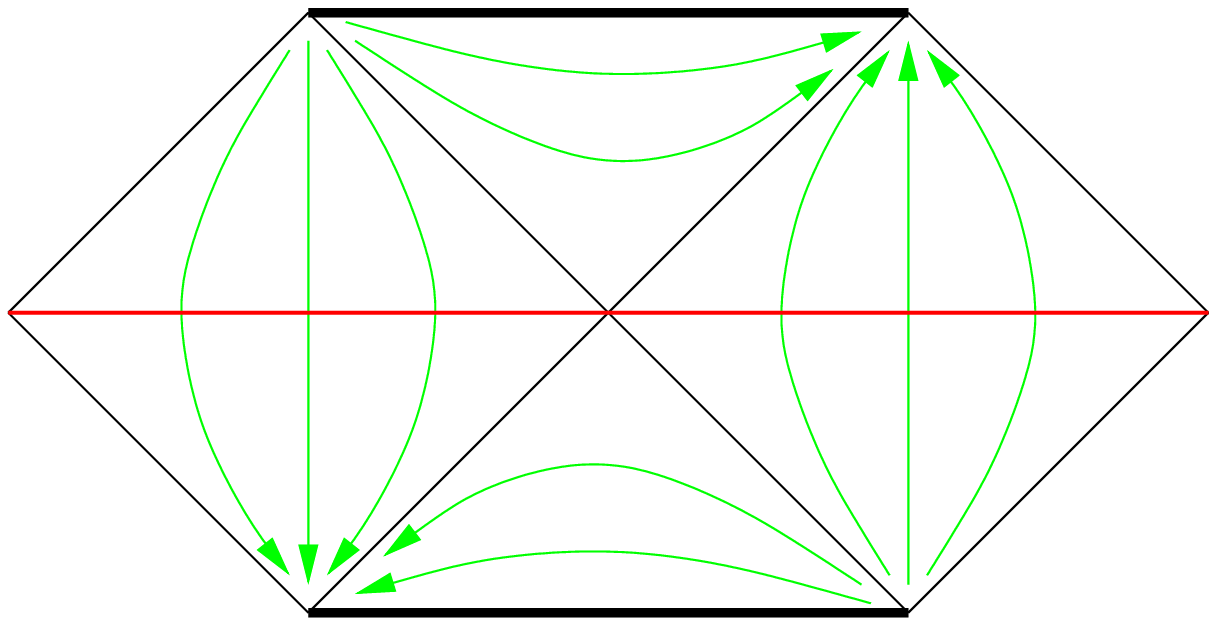}
\caption{\label{fig:kruskalkilling}Orbits of 
Killing time translations on Kruskal. 
In the exteriors these isometries are translations
in Schwarzschild time.\\
$\phantom{xxx}$\\
$\phantom{xxx}$\\
$\phantom{xxx}$}
\end{minipage}\hspace{2pc}%
\begin{minipage}{18pc}
\hspace{4pc}\includegraphics[height=10pc]{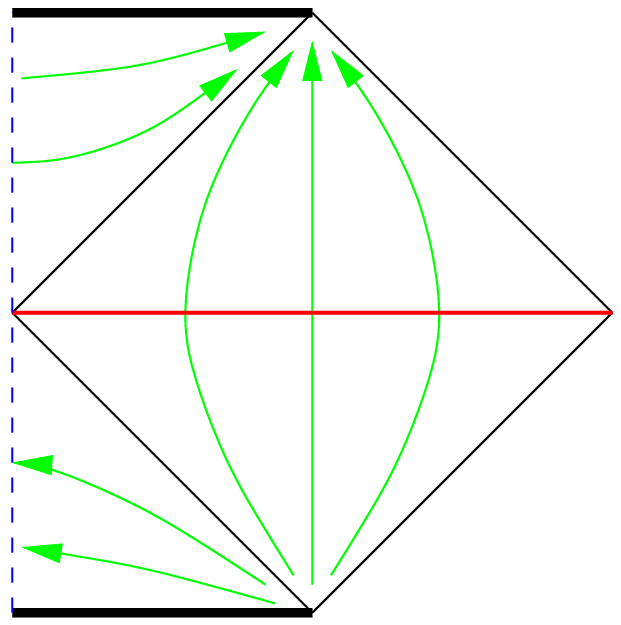}
\caption{\label{fig:rp3geonkilling}Orbits of the local 
$\RPthree$ geon isometry inherited from the 
Killing time translations on Kruskal. 
The isometry is not globally defined because 
the corresponding Killing vector has a sign 
ambiguity in the black and white hole interiors.}
\end{minipage} 
\end{figure}

As is well known, Kruskal admits a one-parameter 
group of isometries that coincide with Schwarzschild 
time translations in the two exteriors. 
These isometries do however not induce 
globally-defined isometries on the $\RPthree$ geon, 
because the quotienting map changes the sign of the 
corresponding Killing vector, as shown in 
Figures \ref{fig:kruskalkilling} and~\ref{fig:rp3geonkilling}. 
As a consequence the geon exterior has a distinguished 
surface of constant Schwarzschild time, 
shown in Figure~\ref{fig:rp3geon}, 
even though one needs to probe the geon geometry 
beyond the exterior to identify this surface. 
The existence of this distinguished spacelike 
surface will turn out significant with 
the Hawking-Unruh effect in Section~\ref{sec:qft}. 

\section{Other geons}
\label{sec:othergeons}

\subsection{Static}

The quotienting of Kruskal into the $\RPthree$ geon generalises 
immediately to the higher-dimensional spherically symmetric 
generalisation of Schwarzschild~\cite{tangherlini}. 
As the antipodal map on a sphere 
preserves the sphere's 
orientation in odd dimensions but reverses 
the sphere's orientation in even dimensions, 
the geon is space orientable 
precisely when the spacetime dimension is even. 

A wider class of generalisations occurs 
when the sphere is replaced by a (not necessarily symmetric) 
Einstein space. 
In $D\ge4$ spacetime 
dimensions, the vacuum solution can be written 
in Schwarzschild-type coordinates as 
\cite{GibbWilt}
\begin{align}
ds^2 = 
-\Delta dt^2 
+ \frac{dr^2}{\Delta}
+ r^2
d{\bar\Omega}_{D-2}^2
\quad , \quad 
\Delta =
{\tilde\lambda} - \frac{
\mu}{r^{D-3}}- 
{\tilde\Lambda} r^2 , 
% \notag 
\label{eq:gibb-wilt-vacuum}
\end{align}
where 
$d{\bar\Omega}_{D-2}^2$ is the metric on the $(D-2)$-dimensional 
Einstein space $\mathcal{M}_{D-2}$, 
${\tilde\lambda}$ is proportional to the 
Ricci scalar of~$\mathcal{M}_{D-2}$, 
${\tilde\Lambda}$ is proportional to the cosmological constant 
and $\mu$ is the mass parameter. 
Suppose that $\Delta$ has a simple zero at 
$r=r_+>0$ and $\Delta>0$ for $r>r_+$. 
The solution has then a Kruskal-type 
extension across a bifurcate Killing horizon at $r=r_+$, 
separating two causally disconnected exteriors. 
If now $\mathcal{M}_{D-2}$ 
admits a freely-acting involutive isometry, a geon quotient 
can be formed just as in Kruskal~\cite{Louko:2004ej}. 
For a negative cosmological constant these geons are 
asymptotically locally anti-de~Sitter. 

For $D=3$ and a negative cosmological constant,  
the above $\BbbZ_2$ quotient 
generalises immediately to the nonrotating 
Ba\~nados-Teitelboim-Zanelli (BTZ) hole \cite{Louko:1998hc}. 
The BTZ hole admits however also geon versions
that arise as other quotients of 
anti-de~Sitter space, even with angular momentum 
\cite{aminneborg-bengtsson-holst,brill-samos,brill-weimar}. 

If a spacelike asymptopia 
at $r\to\infty$ is not assumed, 
further quotient possibilities arise, 
for example by quotienting de~Sitter space or 
Schwarzschild-de~Sitter 
space \cite{louko-schleich,schleich-witt-rp3,McInnes-schwds}. 
We shall not consider these quotients here.

\subsection{Spin}

\begin{figure}[t]
\begin{minipage}{18pc}
\includegraphics[height=10pc]{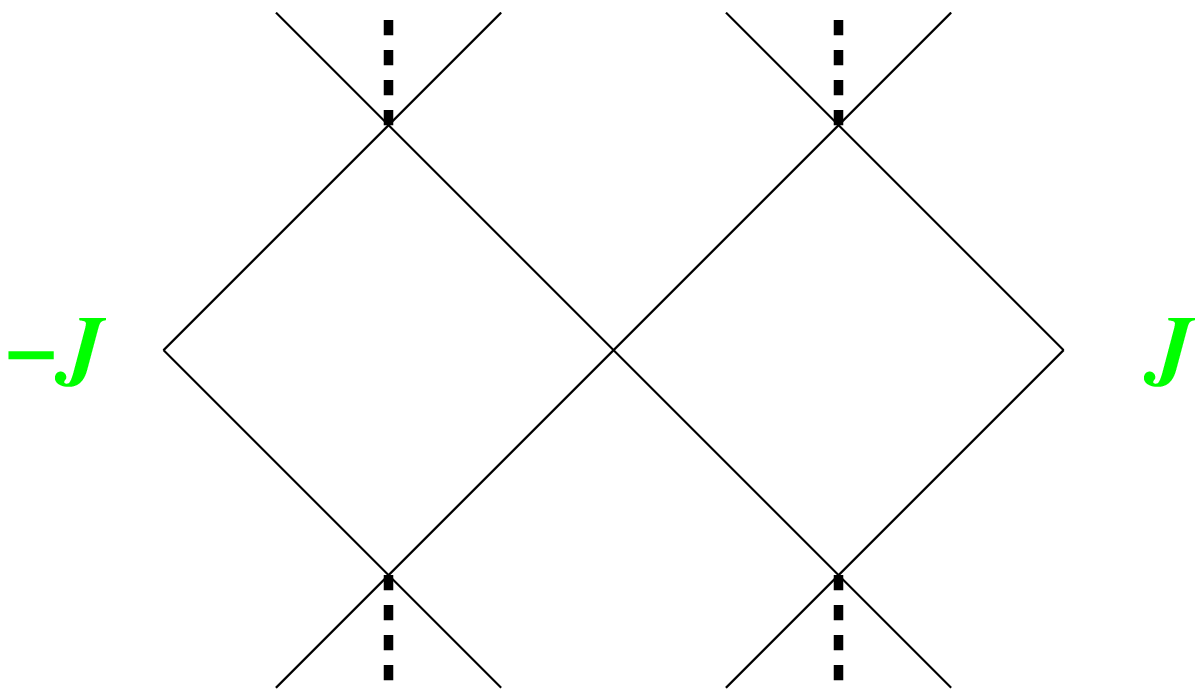}
\caption{\label{fig:kerr}Conformal 
diagram of Kerr. The angular momenta at the 
opposing infinities are equal in absolute 
value but opposite in sign.}
\end{minipage}\hspace{2pc}%
\begin{minipage}{18pc}
\includegraphics[height=10pc]{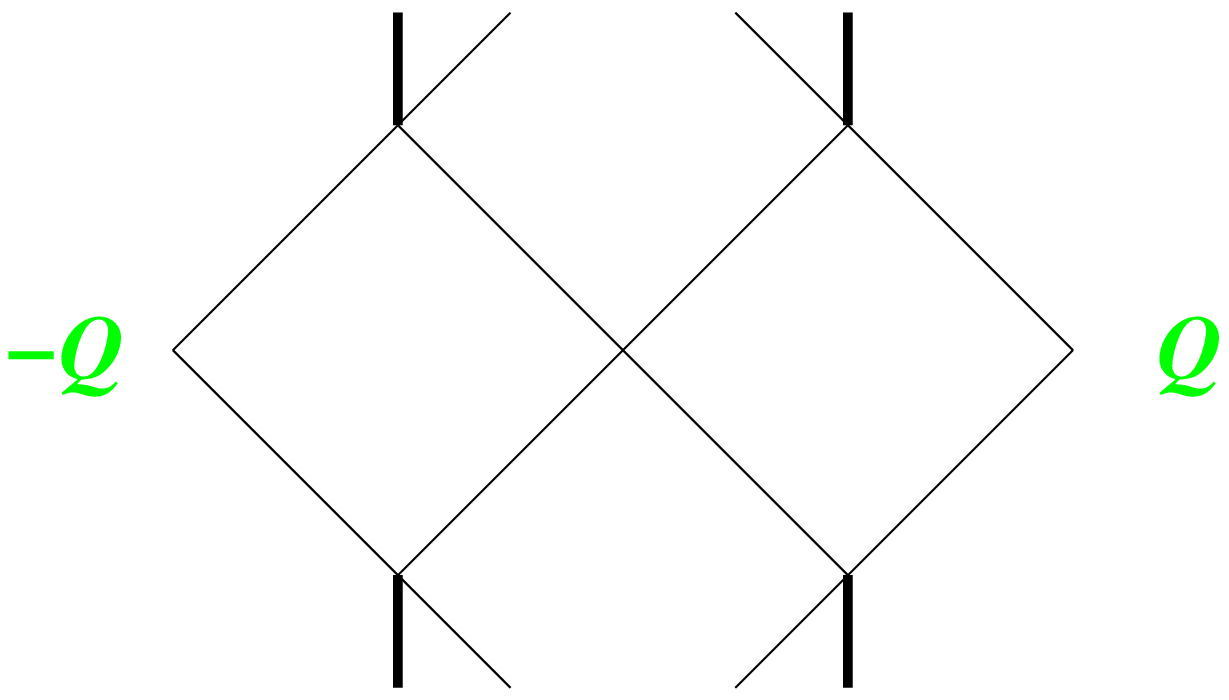}
\caption{\label{fig:reissnord}Conformal 
diagram of Reissner-Nordstr\"om. The electric
charges at the opposing infinities are equal in absolute 
value but opposite in sign.}
\end{minipage} 
\end{figure}

Including spin is delicate. 
To see the problem, consider the Kerr black hole, 
shown in Figure~\ref{fig:kerr}. 
The two exteriors are isometric by an isometry that 
preserves the global time orientation, 
but this isometry inverts the direction
of the rotational angle
and has hence fixed points in the 
black hole region and in the white hole region. 
Kerr does therefore not admit a geon quotient. 
A~way to think about this is that if the 
angular momentum at one infinity is~$J$, 
the angular momentum at the opposing 
infinity is~$-J$. 

The generalisation of Kerr(-AdS) to 
$D\ge5$ dimensions 
\cite{myers-perry,gibbons-lu-page-pope1,%
gibbons-lu-page-pope2,gibbons-perry-pope}
has $[(D-1)/2]$ 
independent angular momenta. 
Each of these angular momenta 
has a different sign at the opposing infinities. 
This suggests that if the angular momenta are arranged into 
multiplets, 
such that within each multiplet the magnitude is constant 
but the signs alternate, 
there might exist a geon quotient by a map that 
suitably permutes the rotation 
planes within each multiplet. 
This turns out to work when $D$ is odd and not equal to~$7$, 
but not for other values of $D$ at least in a way 
that would be continuously deformable 
to the nonrotating case~\cite{Louko:2004ej}. 

For a negative cosmological constant it is 
possible to define angular momenta also on a torus, 
rather than on a sphere. The geon quotient must then again 
involve a permutation of angular momenta with 
equal magnitude and opposite signs~\cite{Louko:2004ej}. 

Finally, the five-dimensional 
Emparan-Reall 
black ring \cite{Emparan:2001wn} 
has a geon quotient~\cite{Chrusciel:2008hg}, despite 
having just one nonvanishing angular momentum. 

\subsection{$\mathrm{U}(1)$ charge}

Including the electromagnetic field is also delicate. 
To see the problem, consider 
electrically-charged Reissner-Nordstr\"om, shown in 
Figure~\ref{fig:reissnord}. 
Given the global time orientation, 
Gauss's law says that if the charge at one 
infinity equals~$Q$, the charge at the opposing 
infinity equals~$-Q$. While there is an involutive 
isometry by which to take a geon quotient 
of the spacetime manifold, 
this isometry reverses the sign the field strength tensor, 
and a quotient is thus not possible 
with the usual understanding of the 
electromagnetic field as a 
$\mathrm{U}(1)$ gauge field~\cite{MW-geondata}. 
The situation is the same 
for the electrically-charged generalisations of 
\eqref{eq:gibb-wilt-vacuum}~\cite{GibbWilt}. 

There is however a way around this minus sign problem. 
Maxwell's theory has a discrete involutive symmetry that 
reverses the sign of the field strength tensor 
as well as the signs of all charges: 
charge conjugation. 
The usual formulation of electromagnetism 
as a $\mathrm{U}(1)$ 
gauge theory treats charge conjugation as a global 
symmetry rather than as a gauge symmetry. 
However, it is possible to promote charge 
conjugation into a gauge symmetry: the gauge group is then no longer 
$\mathrm{U}(1) \simeq \mathrm{SO}(2)$ but 
$\mathrm{O}(2) \simeq \mathbb{Z}_2 \ltimes \mathrm{U}(1)$, 
where in the semidirect product presentation the nontrivial element of 
$\mathbb{Z}_2$ acts on $\mathrm{U}(1)$ 
by complex conjugation~\cite{Kiskis:1978ed}. 
The disconnected component of the enlarged 
gauge group then reverses the sign of the field strength tensor. 
A~geon quotient is now possible provided the quotienting 
map includes a gauge transformation in the disconnected 
component of the enlarged gauge group~\cite{Louko:2004ej}. 
The geon's charge is well-defined only up to the overall sign, 
but detecting the sign ambiguity would 
require observations behind the horizons. 

Within the magnetically-charged 
generalisations of \eqref{eq:gibb-wilt-vacuum}~\cite{GibbWilt}, 
the options depend on the 
$(D-2)$-dimensional 
Einstein space $\mathcal{M}_{D-2}$~\cite{Louko:2004ej}. 
In some cases (including magnetic Reissner-Nordstr\"om 
in even dimensions) the situation is just as with electric charge. 
In other cases (including magnetic Reissner-Nordstr\"om 
in odd dimensions) the field strength tensor 
is invariant under the relevant involutive isometry and 
the usual $\mathrm{U}(1)$ formulation of Maxwell's theory 
suffices. Further options arise if the spacetime has both 
electric and magnetic charge~\cite{Louko:2004ej}. 

\subsection{$\mathrm{SU}(n)$ charge}

Black holes with a non-Abelian gauge field 
also admit a geon 
quotient. 
% ~\cite{Louko:2004ej,kottanattu-thesis,kottanattu-louko}. 
For generic static, spherically symmetric 
$\mathrm{SU}(n)$ black holes with $n>2$ 
\cite{Kuenzle:1994ru,Baxter:2007au,Baxter:2007at} 
this requires 
enlarging the gauge 
group to $\mathbb{Z}_2 \ltimes \mathrm{SU}(n)$, 
where the nontrivial element of $\mathbb{Z}_2$ 
acts on $\mathrm{SU}(n)$ by complex conjugation 
and 
the quotienting map includes a 
a gauge transformation in the disconnected 
component~\cite{kottanattu-thesis,kottanattu-louko}. 
By contrast, static 
$\mathrm{SU}(2)$ black holes
with spherical 
\cite{bizon,kuenzle-masood,win-stability,bjoraker-hoso-small,%
bjoraker-hoso-big,win-sar-even,win-sar-odd}
or axial \cite{Kleihaus:1997ic,Kleihaus:1997ws}
symmetry 
admit a geon quotient without 
enlarging the gauge 
group~\cite{Louko:2004ej,kottanattu-thesis,kottanattu-louko}. 

\section{Quantum field theory}
\label{sec:qft}

The geon black holes that we have discussed are genuine 
black and white holes, 
and the exterior region is isometric to that in an ordinary 
stationary black hole with a bifurcate Killing horizon. 
Does the geon also have a Hawking temperature, 
and if so, in what sense? 

\subsection{Real scalar field on the $\RPthree$ geon}

As the prototype, consider a real 
scalar field on the 
$\RPthree$ geon. 

\begin{figure}[p]
\begin{minipage}{18pc}
\includegraphics[height=12pc]{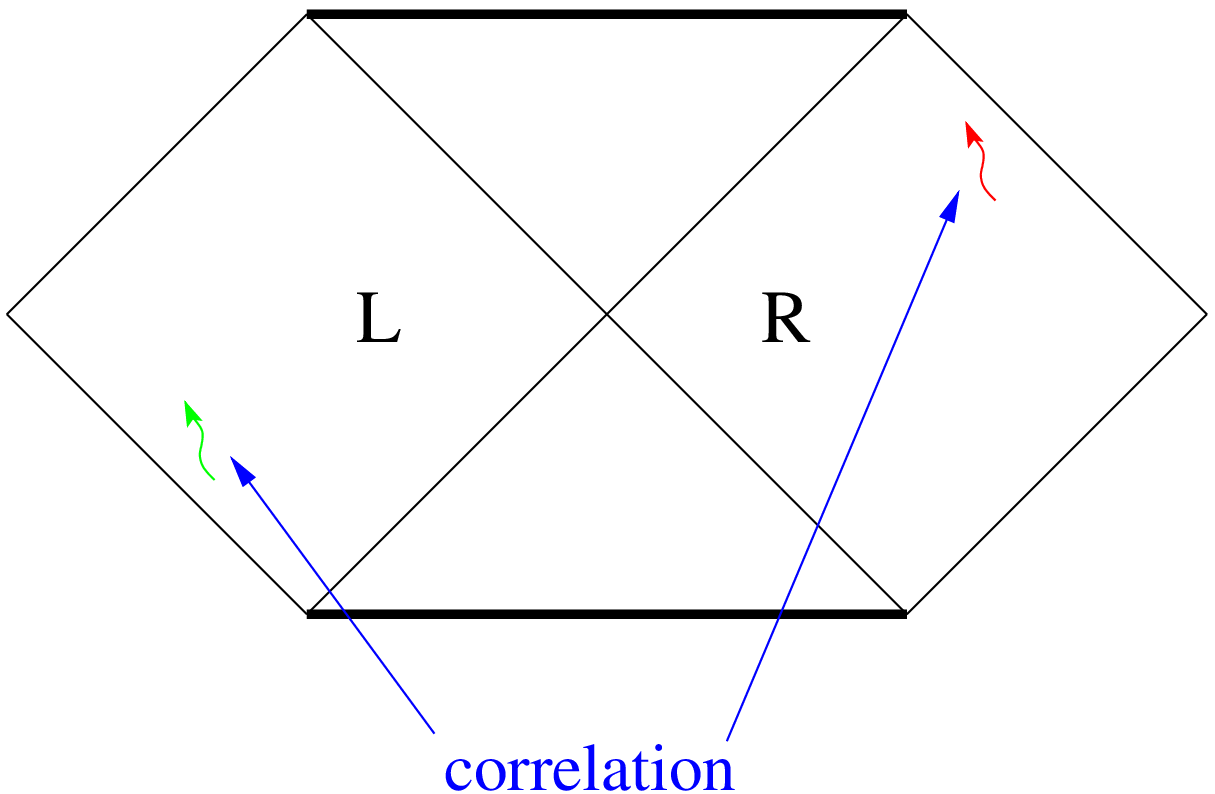}
\caption{\label{fig:kruskalcorr}A correlated pair of 
exterior particles 
in the expansion 
\eqref{eq:vacsexpansion} 
of the Kruskal HHI vacuum $\left|0_{\mathrm{K}}\right\rangle$
as excitations on the 
Boulware vacuum. 
The two particles are in the opposite exteriors. 
From the invariance of $\left|0_{\mathrm{K}}\right\rangle$
under Killing time translations  
it follows that if the particle in $R$ is localised 
in the far future, the correlated particle in $L$ is 
localised in the far past, and vice versa, regardless the precise 
sense of the localisation.}
% An example of localised wave packets 
% is given in~\cite{louko-marolf-rp3}.
\end{minipage}\hspace{2pc}%
\begin{minipage}{18pc}
\hspace{4pc}\includegraphics[height=12pc]{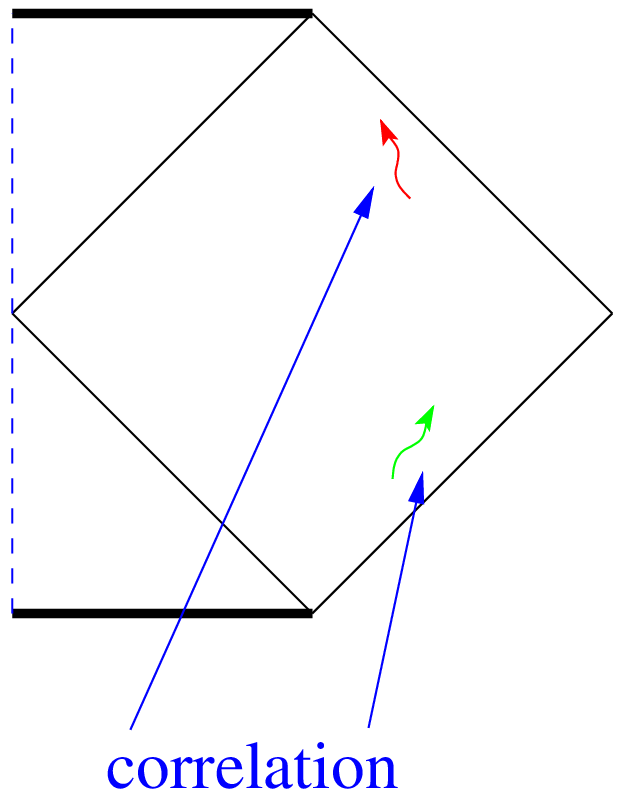}
\caption{\label{fig:rp3geoncorr}A correlated pair of 
exterior particles 
in the expansion 
of the $\RPthree$ geon HHI vacuum
$\left|0_{\mathrm{G}}\right\rangle$
as excitations on the 
Boulware vacuum. 
Both particles are 
in the one and only exterior. 
If one particle is localised 
in the far future, the correlated particle is 
localised in the far past, and vice versa.\\
$\phantom{xxx}$\\
$\phantom{xxx}$}
\end{minipage} 
\end{figure}

Recall first the usual setting for a quantised 
scalar field on Kruskal. 
There is a distinguished vacuum state 
$\left|0_{\mathrm{K}}\right \rangle$, 
the Hartle-Hawking-Israel (HHI) 
vacuum~\cite{Hartle:1976tp,Israel:1976ur}, 
which is the unique regular state
that is invariant under 
all the continuous isometries~\cite{Kay:1988mu}. 
$\left|0_{\mathrm{K}}\right \rangle$ does 
\emph{not\/} coincide with the 
Boulware vacuum state 
$\left|0_{\mathrm{K,B}}\right \rangle$~\cite{Boulware:1974dm}, 
which static observers in the exteriors 
see as a no-particle state. 
Instead, we have the expansion 
\cite{Hartle:1976tp,Israel:1976ur} 
\begin{align}
\left|0_{\mathrm{K}}\right \rangle 
= \sum_{ij\ldots}
f_{ij\ldots}
\underbrace{
a_{R,i}^\dagger
a_{L,i}^\dagger}_{\text{corr}}
\, 
\underbrace{
a_{R,\,j}^\dagger
a_{L,\,j}^\dagger}_{\text{corr}}
\cdots 
\left|0_{\mathrm{K,B}}\right \rangle , 
\label{eq:vacsexpansion}
\end{align}
where 
$a_{R,i}^\dagger$ 
and 
$a_{L,i}^\dagger$ are creation 
operators of particles seen by the static observers in respectively 
the right exterior $R$ and the left exterior~$L$, 
and $i$ stands for the quantum numbers labelling these particles. 
The creation operators 
% in \eqref{eq:vacsexpansion} 
occur in 
correlated pairs, with one particle in each exterior, 
as shown in Figure~\ref{fig:kruskalcorr}. 
From the properties of the expansion coefficients 
$f_{ij\ldots}$ it follows that measurements in (say) $R$ 
see the pure state $\left|0_{\mathrm{K}}\right\rangle$ as a 
Planckian density matrix, 
in a temperature that at asymptotically 
large radii equals the Hawking temperature 
\begin{align}
T_H = \frac{1}{8\pi M} , 
\label{eq:hawkingtemp}
\end{align}
where $M$ is the mass, 
and at finite radii is corrected by the 
gravitational redshift factor. 

To summarise, observers in an exterior region of Kruskal 
see the HHI vacuum 
$\left|0_{\mathrm{K}}\right\rangle$ as a thermal equilibrium state
in the Hawking temperature~\eqref{eq:hawkingtemp}. 

\begin{figure}[p]
\begin{minipage}{18pc}
\includegraphics[height=11pc]{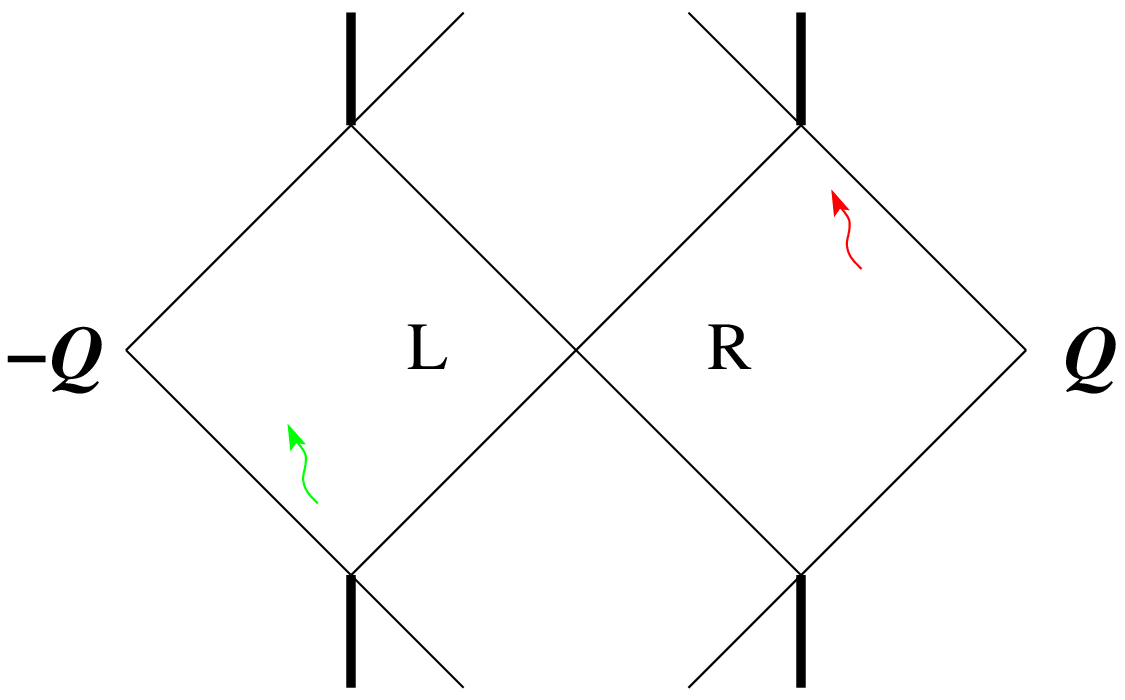}
\caption{\label{fig:reissnordcorr}A correlated pair of 
exterior particles 
in the expansion 
of the electrically-charged Reissner-Nordstr\"om 
HHI vacuum as excitations on the 
Boulware vacuum. 
The two particles have opposite charge.\\
$\phantom{xxx}$\\
$\phantom{xxx}$\\
$\phantom{xxx}$\\
$\phantom{xxx}$}
\end{minipage}\hspace{2pc}%
\begin{minipage}{18pc}
\hspace{5pc}\includegraphics[height=11pc]{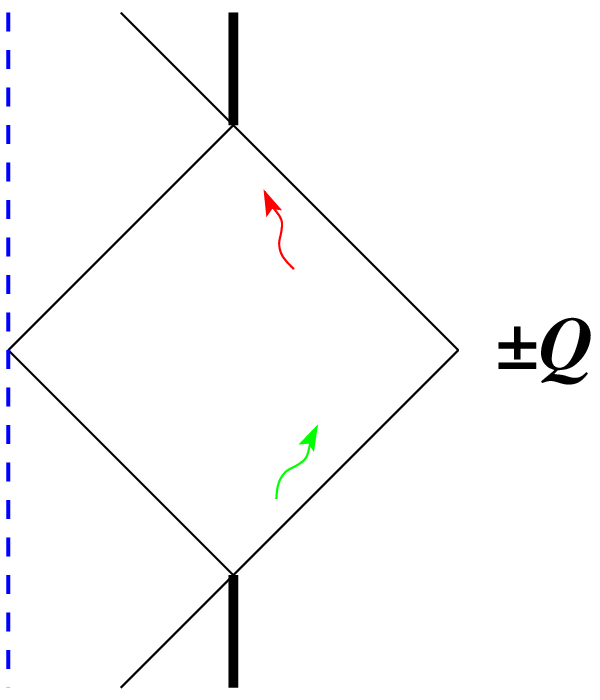}
\caption{\label{fig:reissnordgeoncorr}A correlated pair of 
exterior particles 
in the expansion 
of the electrically-charged 
Reissner-Nordstr\"om 
geon HHI vacuum
as excitations on the 
Boulware vacuum. 
The two particles have the \emph{same\/} charge.\\
$\phantom{xxx}$\\
$\phantom{xxx}$\\
$\phantom{xxx}$\\
$\phantom{xxx}$}
\end{minipage} 
\end{figure}

Consider now a scalar field on the 
$\RPthree$ geon. 
By the quotient from Kruskal, the HHI vacuum 
$\left|0_{\mathrm{K}}\right\rangle$ induces 
on the geon a vacuum state $\left|0_{\mathrm{G}}\right\rangle$, 
which we call the geon vacuum. Again, 
$\left|0_{\mathrm{G}}\right\rangle$
does not coincide with the 
geon's Boulware state $\left|0_{\mathrm{G,B}}\right \rangle$, 
which static observers in the exterior 
see as a no-particle state. 
Instead, 
$\left|0_{\mathrm{G}}\right\rangle$
can be expanded as in~\eqref{eq:vacsexpansion}, 
with creation operators of exterior particles 
acting on $\left|0_{\mathrm{G,B}}\right \rangle$, 
and these creation operators come again in correlated pairs, 
as illustrated in Figure~\ref{fig:rp3geoncorr}. 
For generic observations in the exterior region, 
the geon vacuum state does hence not appear 
as a mixed state, let alone a thermal one. 

However, the correlated particle pairs in the geon 
exterior have a special structure: if one particle 
is localised at asymptotically late times, then the correlated particle 
is localised at asymptotically 
early times, and vice versa. Here, ``early'' and ``late'' 
are defined with respect to the distinguished constant 
Schwarzschild time hypersurface
shown in Figure~\ref{fig:rp3geon}. This means that
$\left|0_{\mathrm{G}}\right\rangle$ appears as a 
mixed state when probed by observers at asymptotically late (or early) 
times, 
and this mixed state 
is again Planckian 
in the Hawking 
temperature~\eqref{eq:hawkingtemp}~\cite{louko-marolf-rp3}. 

To summarise, observers in the geon exterior
see the geon vacuum $\left|0_{\mathrm{G}}\right\rangle$
as thermal in the Hawking temperature \eqref{eq:hawkingtemp} 
at asymptotically late (or early) times. 
At intermediate times the non-thermal correlations in 
$\left|0_{\mathrm{G}}\right\rangle$ 
can be observed by local measurements.

\subsection{Generalisations}

% We now consider some generalisations. 

When the scalar field is replaced by a fermion field, 
a new issue arises in the choice of 
the spin structure~\cite{langlois-rp3}. 
As Kruskal has only one spin structure, 
there is only one HHI vacuum on Kruskal. 
The $\RPthree$ geon has however two spin structures, 
and each of them inherits a 
HHI vacuum from that on Kruskal. 
Observations made at asymptotically late (or early) 
exterior times see each of these vacua as 
a thermal state in the usual Hawking temperature, 
and are in particular unable to distinguish the two vacua. 
The vacua, and their spin structures, 
can however be distinguished by 
observations of the non-thermal 
correlations at intermediate times. 

When the geon has a gauge field, 
a new issue arises from its coupling to 
the quantised matter field. 
For concreteness, consider a complex scalar field on 
electrically-charged Reissner-Nordstr\"om~\cite{bruschi-MGtalk,bruschi-louko}. 
The HHI vacua on the two-exterior hole and on the geon 
again contain exterior particles in correlated 
pairs as shown in Figures 
\ref{fig:reissnordcorr} 
and~\ref{fig:reissnordgeoncorr}, and 
the sense of thermality is as in the uncoupled case but with 
an appropriate chemical potential term~\cite{Gibbons:1975kk}. 
However, while on the two-exterior hole 
the correlated pairs consist of particles 
with opposite charge, 
on the geon the pairs consist of 
particles with the \emph{same\/} charge. 
This is a direct consequence of the charge conjugation
that appears in the quotienting map. 
A geon exterior observer can hence infer 
from the non-thermal correlations 
at intermediate times 
that charges on the spacetime are not globally 
defined, even though this phenomenon only arises 
from the geometry of the geon's gauge 
bundle behind the horizons. 

When the geon has anti-de~Sitter asymptotics, 
the geometry induces a state 
for a quantum field that lives on the geon's 
conformal boundary, via gauge-gravity 
correspondence~\cite{malda-conj,wittensupp}. 
The boundary state exhibits thermality 
for appropriately restricted observations, 
and the non-thermal correlations 
in this state bear an imprint of 
the geometry behind the geon's 
horizons~\cite{Louko:1998hc,lou-ma-ross,malda-eternal,louko-wisniewski}.

\section{Concluding remarks: geon entropy?}
\label{sec:conclusions} 

We have seen that a geon black hole, 
formed as 
a $\BbbZ_2$ quotient of a 
stationary black hole 
with a bifurcate Killing horizon, 
has a Hartle-Hawking-Israel vacuum that appears 
thermal in the usual Hawking temperature 
at asymptotically late (and early) times. 
Does the geon have also an entropy? 

For a late time exterior observer, 
laws of classical black hole mechanics 
do not distinguish between a geon, 
a conventional eternal black hole with a bifurcate Killing horizon, 
or indeed a black hole formed in a star collapse. 
Combining the laws of classical black hole 
mechanics with the late time Hawking temperature, 
the late time observer will hence conclude that 
the geon has the usual Bekenstein-Hawking entropy. 

Could the geon entropy be obtained directly from a quantum theory of gravity? 
If the theory is formulated in a way that allows focussing 
on the quantum spacetime in the distant future, 
state-counting computations for the geon should proceed 
essentially as for the hole with two exteriors. 
This is in particular the case in loop quantum gravity, 
where the counting of black hole microstates is expressly 
set up on the future 
horizon~\cite{Ashtekar:1997yu,Ashtekar:2000eq,Meissner:2004ju,Domagala:2004jt}. 
By contrast, it is not clear whether Euclidean path-integral 
methods allow a localisation at late times. 
While the geon has a regular Euclidean-signature section, 
a na\"\i{}ve application of Euclidean path-integral methods 
to this section yields 
for the entropy a result that 
is only half of the late time Bekenstein-Hawking 
entropy~\cite{louko-marolf-rp3,Louko:1999xb}. 

It would be of interest to develop a geon entropy 
computation within the gauge-gravity correspondence. 
Case studies 
\cite{Louko:1998hc,lou-ma-ross,malda-eternal,louko-wisniewski} 
suggest that an appropriate formalism for this could be late 
time entanglement entropy in the boundary field theory. 
The challenge would be to give a concrete implementation 
in which the gauge theory side of the correspondence is well understood.

\ack

I~am indebted to many colleagues, 
including 
John Barrett, 
David Bruschi, 
John Friedman, 
Gary Gibbons, 
Nico Giulini, 
Bernard Kay, 
George Kottanattu, 
Kirill Krasnov, 
Paul Langlois, 
Robb Mann, 
Don Marolf, 
Simon Ross, 
Kristin Schleich, 
Bernard Whiting 
and Jacek Wi\'sniewski, 
for discussions and collaboration 
on geons. 
I~am grateful to the MCCQG organisers for the 
invitation to participate  
and to present this talk. 
This work was 
supported in part by STFC (UK) grant PP/D507358/1. 
%
%$\phantom{xxx}$ 

%\noindent 
%\emph{This contribution is dedicated to Rafael Sorkin
%in celebration of his influence, 
%inspiration and friendship.} 


\section*{References}
\begin{thebibliography}{99}


\bibitem{wheeler-geon} 
Wheeler J A 
1955 
Geons 
{\it Phys.\ Rev.\ \bf 97} 
511--36 

\bibitem{brill-wheeler-geon} 
Brill D R 
and 
Wheeler J A 
1957
Interaction of
neutrinos and gravitational fields 
{\it Rev.\ Mod.\ Phys.\ \bf 29} 
465--79 

\bibitem{brill-hartle-geon} 
Brill D R 
and 
Hartle J B 
1964
Method of the
self-consistent field in general relativity 
and its application to the
gravitational geon
{\it Phys.\ Rev.\ \bf 135} 
B271--8 

\bibitem{melvin-prd} 
Melvin M A 
1965
Dynamics of cylindrical electromagnetic
universes 
{\it Phys.\ Rev.\ \bf 139} 
B225--43 

\bibitem{MW-geondata} 
Misner C W 
and 
Wheeler J A 
1957
Classical physics as
geometry: Gravitation, electromagnetism, 
unquantized charge, and mass as
properties of curved empty space 
{\it Ann.\ Phys.\ \rm (N.Y.) \bf 2} 
525--603 
Reprinted in 
Wheeler J A 
1962
{\it Geometrodynamics\/} 
(Academic: New York) 

\bibitem{gundlach-rev} 
Gundlach G
2003
Critical phenomena in gravitational
collapse 
{\it Phys.\ Rept.\ \bf 376}
339--405
({\it Preprint\/} gr-qc/0210101)

\bibitem{sorkin-topogeon}
Sorkin R D 
1986
Introduction to topological geons
{\it 
Topological properties and global structure of space-time\/},
Proceedings of the NATO Advanced Study Institute on Topological
Properties and Global Structure of Space-Time, Erice, Italy, May
12--22 1985, 
ed 
P~G Bergmann and V~De~Sabbata
(New York: Plenum) 
pp 249--70 

\bibitem{gannon1} 
Gannon D 
1975
Singularities in nonsimply connected space-times 
{\it J.\ Math.\ Phys.\ \bf 16} 
2364--7 

\bibitem{gannon2} 
Gannon D 
1976
On the topology of spacelike hypersurfaces, 
singularities, and black holes 
{\it Gen.\ Rel.\ Grav.\ \bf 7}
219--32

\bibitem{FriedSchWi} 
Friedman J L, 
Schleich K 
and 
Witt D M 
1993
Topological censorship 
{\it Phys.\ Rev.\ Lett.\ \bf 71} 
1486--9; 
Erratum 
1995
{\it Phys.\ Rev.\ Lett.\ \bf 75} 
1872
({\it Preprint\/} gr-qc/9305017)

\bibitem{Louko:2004ej}
Louko J, 
Mann R B 
and
Marolf D 
2005
Geons with spin and charge 
{\it Class.\ Quant.\ Grav.\  \bf 22}
1451--67
({\it Preprint\/} gr-qc/0412012)

\bibitem{giulini-thesis} 
Giulini D 
1990
3-manifolds in canonical quantum
gravity 
(PhD Thesis, 
University of Cambridge)

\bibitem{chamblin-gibbons} 
Chamblin A
and 
Gibbons G W 
Nucleating black
holes via nonorientable instantons 
{\it Phys.\ Rev.\ \rm D \bf 55}
2177--85
({\it Preprint\/} gr-qc/9607079)

\bibitem{tangherlini} 
Tangherlini F R 
1963
Schwarzschild field in $n$ dimensions 
and the dimensionality of space problem 
{\it Nuovo  Cim.\ \bf 27}
636--51

\bibitem{GibbWilt} 
Gibbons G W
and 
Wiltshire D L 
1987
Space-time as a
membrane in higher dimensions 
{\it Nucl.\ Phys.\ \bf B287}
717--42
({\it Preprint\/} hep-th/0109093)

\bibitem{Louko:1998hc}
Louko J 
and 
Marolf D 
1999 
Single-exterior black holes and the AdS-CFT conjecture 
{\it Phys.\ Rev.\  \rm D \bf 59} 
066002
({\it Preprint\/} hep-th/9808081)

\bibitem{aminneborg-bengtsson-holst} 
{\AA}minneborg S, 
Bengtsson I 
and
Holst S 
1999
A spinning anti-de~Sitter wormhole
{\it Class.\ Quant.\ Grav.\ \bf 16}
363--82
({\it Preprint\/} gr-qc/9805028)

\bibitem{brill-samos} 
Brill D 
2000
Black holes and wormholes in $2+1$
dimensions
{\it Lect.\ Notes Phys.\  \bf 537}
143--79 
({\it Preprint\/} gr-qc/9904083)

\bibitem{brill-weimar} 
Brill D 
2000
$(2+1)$-dimensional black holes with
momentum and angular momentum 
{\it Annalen Phys.\ \rm (Leipzig) \bf 9}
217--26
({\it Preprint\/} gr-qc/9912079)

\bibitem{louko-schleich} 
Louko J 
and 
Schleich K 
1999
The exponential law:
Monopole detectors, Bogolubov transformations, 
and the thermal nature of the
Euclidean vacuum in $\mathbb{RP}^3$ de~Sitter 
spacetime 
{\it Class.\ Quant.\ Grav. \bf 16}
2005--21
({\it Preprint\/} gr-qc/9812056)

\bibitem{schleich-witt-rp3} 
Schleich K 
and 
Witt D M 
1999
The generalized
Hartle-Hawking initial state: 
Quantum field theory on Einstein conifolds
{\it Phys.\ Rev.\ \rm D \bf 60} 
064013
({\it Preprint\/} gr-qc/9903062)

\bibitem{McInnes-schwds} 
McInnes B 
2003
de~Sitter and Schwarzschild-de~Sitter
according to Schwarzschild and de~Sitter  
{\it JHEP\/ \bf 0309} 009
({\it Preprint\/} hep-th/0308022)

\bibitem{myers-perry} 
Myers R C 
and 
Perry M J 
1986
Black holes in
higher-dimensional spacetimes 
{\it Ann.\ Phys.\ \rm (N.Y.) \bf 172}
304--47 

%\bibitem{hawking-hunter-tr} 
%Hawking S W, 
%Hunter C J 
%and 
%Taylor-Robinson M M 
%1999
%Rotation and the AdS/CFT correspondence 
%{\it Phys.\ Rev.\ \rm D \bf 59}
%064005
%({\it Preprint\/} hep-th/9811056)

\bibitem{gibbons-lu-page-pope1} 
Gibbons G W, 
L\"u H, 
Page D N 
and 
Pope C N 
2004
Rotating black holes in higher dimensions with a cosmological constant 
{\it Phys.\ Rev.\ Lett.\ \bf 93} 
171102
({\it Preprint\/} hep-th/0409155)

\bibitem{gibbons-lu-page-pope2} 
Gibbons G W, 
L\"u H, 
Page D N 
and 
Pope C N 
2005 
The general Kerr-de~Sitter metrics in all dimensions 
{\it J. Geom.\ Phys.\ \bf 53} 
49--73
({\it Preprint\/} hep-th/0404008)

\bibitem{gibbons-perry-pope} 
Gibbons G W, 
Perry M J
and 
Pope C N
2005 
The first law of thermodynamics for Kerr-Anti-de~Sitter black
holes
{\it Class.\ Quant.\ Grav.\ \bf 22}
1503--26 
({\it Preprint\/} hep-th/0408217)

\bibitem{Emparan:2001wn}
Emparan R
and 
Reall H S 
2002 
A rotating black ring in five dimensions 
{\it Phys.\ Rev.\ Lett.\  \bf 88} 
101101
({\it Preprint\/} hep-th/0110260)

\bibitem{Chrusciel:2008hg}
Chru\'sciel P T 
and 
Cortier J 
2008 
On the geometry of Emparan-Reall black rings 
{\it Preprint\/} 0807.2309 [gr-qc]

\bibitem{Kiskis:1978ed}
Kiskis J E 
1978
Disconnected gauge groups and the global violation of charge
conservation 
{\it Phys.\ Rev.\  \rm D \bf 17} 
3196--202

\bibitem{Kuenzle:1994ru}
K\"u{}nzle H P 
1994 
Analysis of the static spherically symmetric 
$\mathrm{SU}(n)$ 
Einstein Yang-Mills equations
{\it Commun.\ Math.\ Phys.\  \bf 162} 
371--97

\bibitem{Baxter:2007au}
Baxter J E, 
Helbling M 
and 
Winstanley E 
2007
Soliton and black hole solutions of $\mathrm{SU}(N)$ 
Einstein-Yang-Mills theory in
anti-de~Sitter space
{\it Phys.\ Rev.\  \rm D \bf 76} 
104017
({\it Preprint\/} 0708.2357 [gr-qc])

\bibitem{Baxter:2007at}
Baxter J E, 
Helbling M 
and 
Winstanley E
2008
Abundant stable gauge field hair for black holes 
in anti-de~Sitter space 
{\it Phys.\ Rev.\ Lett.\  \bf 100} 
011301
({\it Preprint\/} 0708.2356 [gr-qc])

\bibitem{kottanattu-thesis}
Kottanattu G T 
2008 
Geon-type black holes in Einstein-Yang-Mills theory 
(PhD thesis, University of Nottingham) 

\bibitem{kottanattu-louko}
Kottanattu G T
and 
Louko J 
(in preparation) 

\bibitem{bizon} 
Bizon P 
1990 
Colored black holes 
{\it Phys.\ Rev.\ Lett.\ \bf 64}
2844--7 

\bibitem{kuenzle-masood} 
K\"unzle H P  
and 
Masood-ul-Alam A K M 
1990
Spherically symmetric static $\mathrm{SU}(2)$ Einstein-Yang-Mills
fields 
{\it J.\ Math.\ Phys.\ \bf 31}
928--35 

\bibitem{win-stability} 
Winstanley E 
1999
Existence of stable hairy black
holes in $\mathrm{SU}(2)$ 
Einstein-Yang-Mills theory with a negative cosmological
constant 
{\it Class.\ Quant.\ Grav.\ \bf 16} 
1963--78
({\it Preprint\/} gr-qc/9812064)

\bibitem{bjoraker-hoso-small} 
Bjoraker J 
and 
Hosotani Y 
2000
Stable monopole
and dyon solutions in the Einstein-Yang-Mills theory in asymptotically
anti-de~Sitter space 
{\it Phys.\ Rev.\ Lett.\ \bf 84} 
1853--6
({\it Preprint\/} gr-qc/9906091)

\bibitem{bjoraker-hoso-big} 
Bjoraker J 
and 
Hosotani Y 
2000
Monopoles, dyons
and black holes in the four-dimensional 
Einstein-Yang-Mills theory
{\it Phys.\ Rev.\ \rm D \bf 62} 
043513
({\it Preprint\/} hep-th/0002098)

\bibitem{win-sar-even} 
Winstanley E 
and 
Sarbach O 
2002
On the linear
stability of solitons and hairy black holes with a negative cosmological
constant: The even parity sector
{\it Class.\ Quant.\ Grav.\ \bf 19}
689--724 
({\it Preprint\/} gr-qc/0111039)

\bibitem{win-sar-odd} 
Winstanley E
and 
Sarbach O 
2001 
On the linear
stability of solitons and hairy black holes with a negative cosmological
constant: The odd parity sector 
{\it Class.\ Quant.\ Grav.\ \bf 18} 
2125--46
({\it Preprint\/} gr-qc/0102033)

\bibitem{Kleihaus:1997ic}
Kleihaus B 
and 
Kunz J
1997
Static black hole solutions with axial symmetry
{\it Phys.\ Rev.\ Lett.\  \bf 79}
1595--8
({\it Preprint\/} gr-qc/9704060)

\bibitem{Kleihaus:1997ws}
Kleihaus B 
and 
Kunz J
1998
Static axially symmetric Einstein-Yang-Mills-dilaton solutions. 
II:  Black hole solutions
{\it Phys.\ Rev.\  \rm D \bf 57}
6138--57 
({\it Preprint\/} gr-qc/9712086)

\bibitem{Hartle:1976tp}
Hartle J B 
and 
Hawking S W 
1976
Path integral derivation of black hole radiance
{\it Phys.\ Rev.\ \rm D \bf 13}
2188--203

\bibitem{Israel:1976ur}
Israel W 
1976
Thermo field dynamics of black holes
{\it Phys.\ Lett.\  \rm A \bf 57}
107--10

\bibitem{Kay:1988mu}
Kay B S 
and 
Wald R M
1991
Theorems on the uniqueness and thermal properties of stationary,
nonsingular, quasifree states on space-times with a bifurcate Killing
horizon
{\it Phys.\ Rept.\  \bf 207}
49--136

\bibitem{Boulware:1974dm}
Boulware D G 
1975
Quantum field theory in Schwarzschild and Rindler spaces
{\it Phys.\ Rev.\  \rm D \bf 11} 
1404--23

\bibitem{louko-marolf-rp3} 
Louko J
and 
Marolf D 
1998
Inextendible
Schwarzschild black hole with a single exterior: 
How thermal is the Hawking
radiation?
{\it Phys.\ Rev.\ \rm D \bf 58} 024007
({\it Preprint\/} gr-qc/9802068)

\bibitem{langlois-rp3} 
Langlois P 
2004
Hawking radiation for Dirac spinors on
the $\mathbb{RP}^3$ geon 
{\it Phys.\ Rev.\ \rm D \bf 70} 104008
({\it Preprint\/} gr-qc/0403011)

\bibitem{bruschi-MGtalk}
Bruschi D E 
2009
Talk given at 
the 
12th Marcel Grossmann meeting, 
Paris, France, 12--18 July 2009. 

\bibitem{bruschi-louko}
Bruschi D E 
and 
Louko J 
(in preparation) 

\bibitem{Gibbons:1975kk}
Gibbons G W 
1975
Vacuum polarization and the spontaneous loss of charge by black holes
{\it Commun.\ Math.\ Phys.\  \bf 44}
245--64

\bibitem{malda-conj}
Maldacena J 
1998
The large $N$ limit of superconformal field theories and
supergravity 
{\it Adv.\ Theor.\ Math.\ Phys.\ \bf 2}
231--52 
({\it Preprint\/} hep-th/9711200)

\bibitem{wittensupp}
Witten E 
1998
Anti-de~Sitter space and holography 
{\it Adv.\ Theor.\ Math.\ Phys.\ \bf 2}
253--91 
({\it Preprint\/} hep-th/9802150)

\bibitem{lou-ma-ross}  
Louko J, 
Marolf D 
and 
Ross S F 
2000
Geodesic propagators and black hole holography
{\it Phys.\ Rev.\ \rm D \bf 62} 044041
({\it Preprint\/} hep-th/0002111)

\bibitem{malda-eternal}
Maldacena J M 
2003
Eternal black holes in anti-de~Sitter 
{\it JHEP\/ \bf 0304} 021
({\it Preprint\/} hep-th/0106112)

\bibitem{louko-wisniewski} 
Louko J 
and 
Wi\'sniewski J 
2004
Einstein black holes, free scalars and 
AdS/CFT correspondence 
{\it Phys.\ Rev.\ \rm D \bf 70} 084024
({\it Preprint\/} hep-th/0406140)

\bibitem{Ashtekar:1997yu}
Ashtekar A, 
Baez J, 
Corichi A 
and 
Krasnov K 
1998
Quantum geometry and black hole entropy 
{\it Phys.\ Rev.\ Lett.\  \bf 80} 
904--7
({\it Preprint\/} gr-qc/9710007)

\bibitem{Ashtekar:2000eq}
Ashtekar A, 
Baez J 
and 
Krasnov K 
2000
Quantum geometry of isolated horizons and black hole entropy 
{\it Adv.\ Theor.\ Math.\ Phys.\  \bf 4}
1--94
({\it Preprint\/} gr-qc/0005126)

\bibitem{Meissner:2004ju}
Meissner K A 
2004
Black hole entropy in loop quantum gravity 
{\it Class.\ Quant.\ Grav.\  \bf 21}
5245--52
({\it Preprint\/} gr-qc/0407052)

\bibitem{Domagala:2004jt}
Domagala M 
and 
Lewandowski J 
2004
Black hole entropy from quantum geometry 
{\it Class.\ Quant.\ Grav.\  \bf 21}
5233--44
({\it Preprint\/} gr-qc/0407051)

\bibitem{Louko:1999xb}
Louko J 
2000
Single-exterior black holes 
{\it Lect.\ Notes Phys.\  \bf 541}
188--202
({\it Preprint\/} gr-qc/9906031)

\end{thebibliography}
\end{document}